\documentclass[
reprint,
superscriptaddress,
groupedaddress,
longbibliography,
bibnotes,
amsmath,amssymb,
aps,
prb,
]{revtex4-2}

\usepackage[utf8]{inputenc}
\usepackage[english]{babel}
\usepackage[plainpages = false, pdfpagelabels, 
                 bookmarks,
                 bookmarksopen = true,
                 bookmarksnumbered = true,
                 breaklinks = true,
                 linktocpage,
                 colorlinks = true,
                 linkcolor = blue,
                 urlcolor  = blue,
                 citecolor = blue,
                 anchorcolor = green,
                 hyperindex = true,
                 hyperfigures
                 ]{hyperref} 
\usepackage{enumerate}
\usepackage{tabularx}
\usepackage{makecell, multirow}
\usepackage{tikz}
\usetikzlibrary{patterns}
\usepackage{slashed}
\usepackage{physics}
\usepackage{verbatim}
\usepackage{cancel}
\usepackage{bbold}
\usepackage{bm}
\usepackage{booktabs}
\usepackage[caption=false,singlelinecheck=off]{subfig}
\usepackage{blindtext}
\usepackage{bbm}
\usepackage{graphicx}
\usepackage{dcolumn}
\usepackage{bm}
\usepackage[normalem]{ulem}

\usepackage{xcolor}

\newcommand*{\Sth}[1][I]{S_\text{th}^{#1}}
\newcommand*{\Ssh}[1][I]{S_\text{sh}^{#1}}
\newcommand*{\Sr}[1][I]{S_\text{rest}^{#1}}
\renewcommand*{\L}{\text{L}}
\newcommand*{\R}{\text{R}}
\newcommand*{\fL}{f_\L(E)}
\newcommand*{\fR}{f_\R(E)}
\newcommand*{\Fal}{f_\alpha(E+\hbar\omega)}
\renewcommand*{\FL}{f_\L(E+\hbar\omega)}
\renewcommand*{\FR}{f_\R(E+\hbar\omega)}
\newcommand*{\fa}{f_\alpha(E)}
\newcommand*{\fat}{f_{\alpha\tau}(E)}

\newcommand*{\kB}{k_\text{B}}
\newcommand*{\cT}{\Theta}
\newcommand{\cR}{\mathcal{R}}
\newcommand{\cA}{\mathcal{A}}

\newcommand{\subfigref}[2]{\ref{#1}\hyperref[#1]{(#2)}}
\newcommand{\Li}{\mathrm{Li}}

\begin{document}

\title{\texorpdfstring{Charge, spin, and heat shot noises in the absence of average currents:\\ Conditions on bounds at zero and finite frequencies}{}}
\author{Ludovico Tesser}
\author{Matteo Acciai}
\author{Christian Sp\r{a}nsl\"{a}tt}
\author{Juliette Monsel}
\author{Janine Splettstoesser}
 \affiliation{Department of Microtechnology and Nanoscience (MC2),Chalmers University of Technology, S-412 96 G\"oteborg, Sweden}
\date{\today}
\begin{abstract}
Nonequilibrium situations where selected currents are suppressed are of interest in fields like thermoelectrics and spintronics, raising the question of how the related noises behave.
We study such zero-current charge, spin, and heat noises in a two-terminal mesoscopic conductor. 
In the presence of voltage, spin and temperature biases, the nonequilibrium (shot) noises of charge, spin, and heat can be arbitrarily large, even if their average currents vanish. However, as soon as a temperature bias is present, additional equilibrium (thermal-like) noise necessarily occurs. We show that this equilibrium noise sets an upper bound on the zero-current charge and spin shot noises, even if additional voltage or spin biases are present. We demonstrate that these bounds can be overcome for heat transport by breaking the spin and electron-hole symmetries, respectively. By contrast, we show that the bound on the charge noise for strictly two-terminal conductors even extends into the finite-frequency regime.
\end{abstract}
\maketitle

\section{\label{sec:Introduction}Introduction}
Fluctuations, or noise, in physical observables disclose important properties of small electronic conductors. While equilibrium charge noise relates a conductor's temperature to its dc conductance according to the Nyquist-Johnson relation~\cite{Nyquist1928,Johnsson1928}, nonequilibrium noise offers additional opportunities. Most prominently, shot noise---or partition noise---which arises from the granularity of the electric charge, has in the last decades emerged as a ubiquitous tool for characterizing nanoscale systems~\cite{Schottky1918,BlanterButtikerPhysRep00,Kobayashi2021}. It has, e.g., been used to reveal the charge of fractionalized quasiparticles~\cite{Saminadayer1997,De-Picciotto1997Sep}, Cooper pairs~\cite{Kozhevnikov2000Apr,Jehl2000May} as well as Bogoliubov quasiparticles~\cite{Ronen2016} in superconductors. Besides, analyzing and understanding  nonequilibrium noise in  nanoscale thermoelectric devices is crucial as it limits their performances~\cite{Seifert2018Aug,Pietzonka2018May,Kheradsoud2019Aug,Pal2020May,Crepieux2014Dec,Eymeoud2016Nov,Crepieux2016May,Sayral2021Nov,Gerry2022Feb}.

\begin{figure}[b!]
    \centering
    \includegraphics{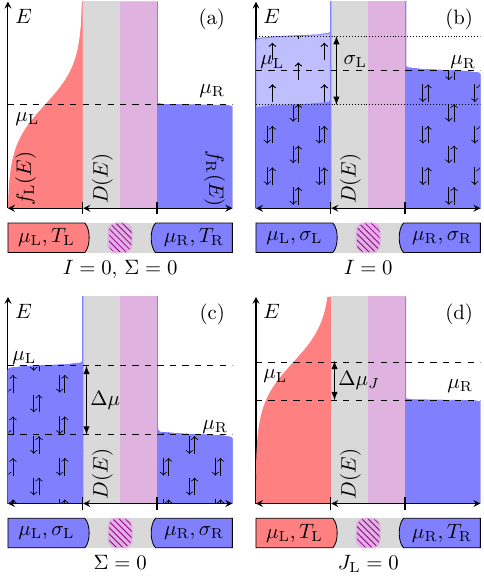}
    \caption{Zero-current conditions  obtained for charge ($I$), spin ($\Sigma$), and heat ($J_\L$) currents for the case of uniform transmission $D(E)=1/2$ (depicted in purple).
    (a) Using only a temperature bias, both average charge and spin currents vanish.
    (b) The zero-charge-current condition is achieved by using only a spin imbalance $\sigma_\L$ on the left contact.
    (c) The zero-spin-current condition is achieved by using only a voltage bias $\Delta\mu$ between the contacts.
    (d) The zero-heat-current condition, $J_\L=0$, is achieved using both a voltage and a temperature bias. The voltage bias can be replaced by a spin imbalance.}
    \label{fig:D-uniform}
\end{figure}

More recently, charge noise as a response to a temperature gradient and in the absence of a voltage bias---dubbed delta-$T$ noise---has been investigated in several theoretical~\cite{Sukhorukov1999,Mu2019,Zhitlukhina2020,RechMartinPRL20,Hasegawa2021,duprezX2021,Popoff2022,Eriksson2021Sep,Schiller2022,Zhang2022,Rebora2022Jul} and experimental~\cite{TikhonovSciRep16,Lumbroso2018,Sivre2019,TikhonovPRB2020,Larocque2020,Rosenblatt2020,Melcer2022} studies.
These studies demonstrate that delta-$T$ noise offers additional insights beyond the traditional shot noise, e.g., for quantifying local temperature gradients~\cite{Lumbroso2018,duprezX2021} or as a potential tool for extracting scaling dimensions and exchange statistics of anyons~\cite{RechMartinPRL20,Schiller2022,Zhang2022}. Still, the full scope of delta-$T$ noise remains to be understood. 

A key feature of delta-$T$ noise is that, for energy independent transport through the system, the absence of voltage bias causes the average charge current to vanish. Yet, the conductor partitions the opposing, equal current contributions emanating from the reservoirs, which results in detectable shot noise. It was, however, pointed out in Ref.~\cite{Eriksson2021Sep} that noise in the absence of a current is a broader concept than delta-$T$ noise.
More specifically, one can imagine experimental setups with generic transmissions, where temperature and voltage biases are carefully combined such that the average current vanishes. This is particularly relevant for characterizing thermoelectric properties of nanodevices: A zero charge current situation indeed occurs at the stopping voltage, also referred to as the thermovoltage. Under open circuit conditions, the zero-current condition is exactly that at which a device's thermopower---the ratio between the thermovoltage and the temperature bias---is extracted. Moreover, zero-current nonequilibrium noise is not limited to charge transport, but can also be considered for currents of, e.g., heat~\cite{Pekola2021Oct} or spins~\cite{Hirohata2020Sep}, see Fig.~\ref{fig:D-uniform}.

In this paper, we focus on zero-current fluctuations and compare how nonequilibrium fluctuations behave compared with their equilibrium-like counterpart. We extend the analysis of Ref.~\cite{Eriksson2021Sep}, for a coherent, mesoscopic conductor, characterized by a transmission function $D(E)$, which is connected to two macroscopic reservoirs (equivalently terminals, or contacts). Within a scattering approach~\cite{BlanterButtikerPhysRep00}, we compute the noise of charge, heat, and spin currents by combining external biases and transmission functions such that one or more average currents vanish. More precisely, we always consider a vanishing average current in a given contact for the same transported quantity as for the studied noise, e.g., heat shot noise with zero average heat current in one contact.

First, we broaden the scope of delta-$T$ noise by showing that, even in the simple case of constant transmission $D(E)=D$, charge, spin and heat shot noises can arise under zero-current conditions and have similar functional forms. These zero-current shot noises can become arbitrarily large with increasing magnitude of the bias causing the nonequilibrium situation. Furthermore, we discuss how the nonequilibrium noise is related to the flow of excitations between the reservoirs, under the condition that the applied biases are large with respect to the energy scale of the base temperature.

Next, when a temperature is comparable to the energy scales set by the biases, it makes sense to ask how large the nonequilibrium fluctuations can be with respect to the thermal component of the noise. We tackle this problem for an arbitrary transmission function $D(E)$. For a spin-degenerate system, it was shown in Ref.~\cite{Eriksson2021Sep} that the zero-current charge shot noise is always bounded by its thermal counterpart, independently of the conductor's transmission function.
Here, we extend this result in several ways:

(i) We obtain an even stricter upper bound, given in Eq.~\eqref{eq:bound-shot-charge}, than the one previously derived in Ref.~\cite{Eriksson2021Sep}, see Eq.~\eqref{eq:Charge_shot_0}, which is valid at arbitrary reservoir temperatures $T_\L, T_\R$ and transmission function $D(E)$.
 
(ii) We show that the new bound~\eqref{eq:bound-shot-charge} also applies to the zero-current spin noise in the presence of spin and temperature biases.

(iii) We demonstrate that the zero-current charge noise remains bounded at finite frequency [see Eq.~\eqref{eq:finite-freq-bound}], provided the noise is measured in the colder reservoir. By contrast, if the noise is measured in the hotter reservoir, the bound~\eqref{eq:finite-freq-bound} does not hold. 

Our findings in this paper highlight several important features of zero-current nonequilibrium noise underlining that it could be used as a future noise spectroscopic tool, in particular, to probe nanoscale gradients of temperature~\cite{Lumbroso2018,duprezX2021} or spin polarization.

The remainder of this paper is structured as follows. In Sec.~\ref{sec:Model}, we introduce the here employed scattering-based formalism.
In Sec.~\ref{sec:constant-transmission} we extend the concept of delta-$T$ noise to different kinds of bias and apply it to charge, spin, and heat currents.
In Sec.~\ref{sec:shot-thermal-noise}, we demonstrate how to achieve unbounded zero-current nonequilibrium heat noise, and present an improved bound for the charge noise, which also holds for the spin noise.
The bound on charge shot noise is further extended to the finite-frequency noise in Sec.~\ref{sec:finite-frequency}. In Sec.~\ref{sec:measurement_scheme}, we address the experimental prospects to verify our bounds for the zero-current charge noise.

\section{\label{sec:Model}Scattering approach to noise}
We study steady-state transport in a coherent quantum conductor connected to two macroscopic reservoirs, labeled by $\alpha=\L,\R\,$. 
The conductor is characterized by a spin-independent transmission function $D(E) = |s_\text{LR}(E)|^2 = 1-|s_\text{LL}(E)|^2$, obtained from a spin-preserving scattering matrix
\begin{equation}
    s(E) = \begin{pmatrix} s_\text{LL}(E) & s_\text{LR}(E)\\ s_\text{RL}(E) & s_\text{RR}(E)\end{pmatrix}.
\label{eq:smatrix}
\end{equation}
The electronic occupations in the reservoirs are governed by Fermi distribution functions
\begin{align}
\label{eq:Fermi_functions}
    f_{\alpha\tau}(E) = \frac{1}{1+e^{\beta_\alpha(E -\mu_{\alpha\tau})}},
\end{align}
where $\beta_\alpha=(\kB T_\alpha)^{-1}$ are the inverse temperature scales, and $\kB$ is the Boltzmann constant. When the spin degeneracy for $\tau=\uparrow,\downarrow$ in the reservoirs is broken, we write the spin-dependent electrochemical potentials as
\begin{equation}
    \label{eq:chemical_spin}
    \mu_{\alpha\tau}=\mu_{\alpha}-(-1)^{\delta_{\tau\downarrow}}\frac{\sigma_\alpha}{2},
\end{equation}
where $\delta_{\tau\tau'}$ is the Kronecker delta and the spin-splitting in reservoir $\alpha$ is given by $\sigma_\alpha$. We are interested in nonequilibrium situations, where the distributions of the two reservoirs differ due to any of the three biases $\Delta\mu=\mu_\text{L}-\mu_\text{R}$, $\Delta T=T_\text{L}-T_\text{R}$, or $\Delta\sigma=\sigma_\text{L}-\sigma_\text{R}$.

The average charge ($I$), heat ($J$) and spin ($\Sigma$) currents, which can possibly flow out of the left contact in response to these three biases and their combinations, are given by~\cite{BlanterButtikerPhysRep00}
\begin{align}\label{eq:current-definition}
   X_\L=\frac{1}{h}{\sum_\tau}\int_{{-\infty}}^{{\infty}} dE\,xD(E)[{f_{\L\tau}(E)-f_{\R\tau}(E)}]\,,
\end{align}
with $x\to\{-e,E-\mu_{\L\tau},{(-1)^{\delta_{\tau\downarrow}}\hbar/2}\}$ for $X\to\{I,J,{\Sigma}\}$ and analogously for $X_\R$.
Here, $e>0$ is the elementary charge (the electron charge is thus $-e$), and $h\equiv 2\pi\hbar$ is the Planck constant. All energies are measured with respect to a reference electrochemical potential (e.g., $\mu_0=(\mu_\R+\mu_\L)/2$) and, unless otherwise specified, all energy integrals in the remainder of the paper are to be understood as $\int dE=\int_{-\infty}^{\infty} dE$.

We define the noise at frequency $\omega$ associated with the current $X$ as~\cite{BlanterButtikerPhysRep00}
\begin{align}
\label{eq:Noise_def}
    S^X_{\alpha\beta}(\omega)=\int_{{-\infty}}^{{\infty}}\langle\{\delta \hat{X}_\alpha(t),\delta \hat{X}_\beta(0)\}\rangle e^{i\omega t}dt,
\end{align}
where $\delta \hat{X}_\alpha=\hat{X}_\alpha-X_\alpha$ is the fluctuation of the operator $\hat{X}_\alpha$ around its thermal average value $X_\alpha\equiv \langle \hat{X}_\alpha \rangle$, and $\{\hat{X}_\alpha,\hat{Y}_\beta \} = \hat{X}_\alpha \hat{Y}_\beta+\hat{Y}_\beta\hat{X}_\alpha$ is the anticommutator.
In the following, we study the left autocorrelator $S^X(\omega)\equiv S^X_{\L\L}(\omega)$, which corresponds to measuring the current fluctuations in the left contact.
In the first part of the paper, we will focus on the zero-frequency regime, $\omega=0$, and analyze the noise for various types of biases and associated currents. Here, the noise of a conserved current, in our case both charge and spin current $X\to\{I, \Sigma\}$, satisfies $S_{\L\L}^X(0)=S_{\R\R}^X(0)=-S_{\L\R}^X(0)=-S_{\R\L}^X(0)$.
By contrast, these conservation laws do not hold for heat noise, nor for noise at finite frequency~\cite{BlanterButtikerPhysRep00}; in these specific cases, the noise depends on the contact in which it is measured.
The analysis of finite-frequency noise is reported in Sec.~\ref{sec:finite-frequency} and focuses on the charge noise.

At $\omega=0$, the noise can be written in a compact form and is straightforwardly separated into two contributions, $S^X(0)=\Sth[X](0)+\Ssh[X](0)$, with~\cite{BlanterButtikerPhysRep00,Moskalets2011Sep}
\begin{subequations}
\label{eq:noise-general}
\begin{align}
    \Sth[X](0)&=\sum_{\alpha\tau} \!\int\! dE \frac{2x^2}{h} D(E)f_{\alpha\tau}(E)[1-f_{\alpha\tau}(E)]\,,
    \label{eq:thermal-general}\\
    \Ssh[X](0)&=\sum_{\tau} \!\int\! dE \frac{2x^2}{h} D(E)[1-D(E)][f_{\L\tau}(E)-f_{\R\tau}(E)]^2.
\label{eq:shot-general}
\end{align}   
\end{subequations}
Here, $\Sth[X](0)$ is thermal-like noise to which each reservoir contributes individually, even at equilibrium, $f_{\L\tau}=f_{\R\tau}$. By contrast, $\Ssh[X](0)$ is the so-called shot noise which is nonzero only under nonequilibrium conditions, i.e., when $f_{\L\tau}\neq f_{\R\tau}$. It contains the characteristic partitioning factor $D(E)[1-D(E)]$
and thus vanishes in the limits of perfect, $D(E)=1$, or completely suppressed transmission, $D(E)=0$. Note that the factors of 2 in Eq.~\eqref{eq:noise-general} come from the anticommutator in Eq.~\eqref{eq:Noise_def}.

Of main interest for our work is the \emph{shot noise}, Eq.~\eqref{eq:shot-general}, under the \emph{zero-current condition} 
\begin{align}
\label{eq:zero_current_condition}
X_\L = 0.    
\end{align}
Depending on the transmission function $D(E)$ and on the type of current that should vanish, a combination of biases $\Delta\mu,\Delta T$ and $\Delta\sigma$ is required. Note that for a conserved current (charge and spin) $X_\L=0\implies X_\R=0$.  This conservation does not hold for the heat current, which can be made to vanish only in one contact at a time. Here, we impose $X_\L=0$, consistent with our choice to study the noise correlator in the left contact $S^X(\omega)\equiv S^{X}_{\L\L}(\omega)$.

\section{Charge, spin, and heat noises at zero average current}\label{sec:constant-transmission}
We begin by presenting charge, spin, and heat fluctuations in the absence of the related average currents under nonequilibrium conditions determined by different kinds of biases, see Fig.~\ref{fig:D-uniform}.
We focus here on the situation where the possible applied biases---the potential bias $\Delta\mu$, temperature bias $\Delta T$, or spin bias $\Delta \sigma$---are large, resulting in large shot noise. Concretely, this means for any of the applied biases, $\Delta\mu,\kB\Delta T,\Delta\sigma\gg \kB T_\text{R}$, such that we can effectively set $T_\text{R}\to 0$ and $\Delta T=T_\text{L}$ in the present section. In Appendix~\ref{app:smallbias}, we present results for the zero-current heat noise in the opposite regime of weak biases, thus complementing the literature for the zero-current charge shot noise in this regime~\cite{Lumbroso2018}.

To start with, we consider a uniform transmission function, $D(E)=D$. This simple choice results in electron-hole as well as spin symmetry in the scattering process. Consequently, shot noise at vanishing average charge or spin currents can be obtained in the presence of a single type of bias. Concretely, zero current is here obtained when the applied bias does not break the symmetry related to the transported observable: temperature and spin biases do not break electron-hole symmetry, resulting in zero charge current; temperature and voltage biases do not break the spin symmetry, resulting in zero spin current. We show that in these cases, current cancellation results from incoming fluxes of opposite sign, which, however, sum up to a nonvanishing contribution to the nonequilibrium (shot) noise. 
In contrast, to reach the zero-heat-current condition at constant transmission, it is necessary to have at least two biases, one of which being the temperature bias. The reason for this is that any of the three biases breaks the symmetry with respect to the excess energy transported into the left contact and needs hence to be (nontrivially) compensated by a second bias. Only when the transmission has an appropriate energy-dependence can a zero heat current be reached by the application of a single bias.
All different possible settings, which we present in this section, are listed in Tab.~\ref{tab:table}.

\begin{table}[tb]
\centering
\includegraphics{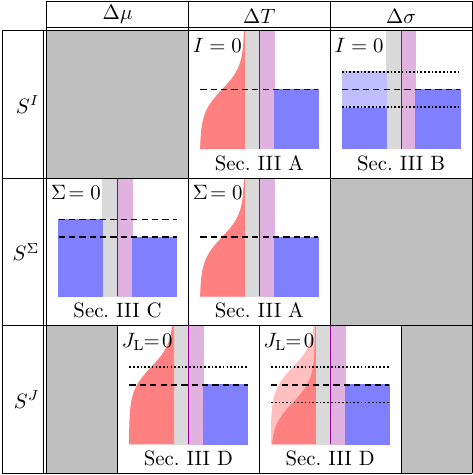}
\caption{Realizations of generalized zero-current shot noise, for energy-independent transmissions $D(E)=D$ (purple).}\label{tab:table}
\end{table}

\subsection{\texorpdfstring{Delta-$T$ charge and spin current noises}{}}\label{subsec:delta-T-noise}
First, we consider a spin-degenerate setup in which only a temperature bias between the contacts generates the nonequilibrium noise, as illustrated in Fig.~\subfigref{fig:D-uniform}{a}.
This is the situation in which the delta-$T$ noise was studied in Refs.~\cite{Lumbroso2018,Sivre2019,Larocque2020}.
In this case, transport can be understood in terms of negatively (electrons) or positively (holes) charged excitations flowing from the hot contact to the cold one. Therefore, both the average charge and spin currents vanish
\begin{subequations}
    \begin{eqnarray}
I       & = & -e\sum_{\tau=\uparrow,\downarrow} \left(j_{\text{e}\tau}-j_{\text{h}\tau}\right)  =0,\label{eq:I0}\\
\Sigma  & = &\frac{\hbar}2\sum_{i=\text{e,h}} \left(j_{i\uparrow}-j_{i\downarrow}\right)        =0.\label{eq:Sigma0}
\end{eqnarray}
\end{subequations}
Here, we have defined the influxes of spin-resolved excitations from the left contact as 
\begin{subequations}
\begin{align}\label{eq:eh-flow}
    j_{\text{e}\tau} &= \frac{D}{h}\int_{\mu_\text{R}}^\infty\!dEf_{\L\tau}(E), \\ 
    j_{\text{h}\tau} & =\frac{D}{h} \!\int_{-\infty}^{\mu_\text{R}}\!dE[1-f_{\L\bar{\tau}}(E)],
\end{align}
\end{subequations}
where the notation $\bar{\tau}$ refers to the opposite spin of $\tau$. In Eq.~\eqref{eq:I0}, we classify the excitations according to their charge, while we classify them according to their spin in Eq.~\eqref{eq:Sigma0}. 

Unlike the currents, the noise is finite, and importantly it contains a nonvanishing \textit{shot} noise component
\begin{align}
    S^I(0)& = S^I_\text{sh}(0)+S^I_\text{th}(0)\\
    & = k_\text{B} T_\text{L}\frac{4e^2}{h}D(1-D)(2\ln 2-1) + k_\text{B} T_\text{L}\frac{4e^2}{h}D.\nonumber
\end{align}
When the conductor is opaque, namely $D\ll 1$, the charge-current fluctuations are given by
\begin{equation}\label{eq:delta-T-noise}
    S^I(0) \approx \frac{8e^2}{h} D k_\text{B} T_\text{L}\ln2 = 2e^2 j.
\end{equation}
Here, we recognize that the charge current noise is proportional to the influx of excitations $j$ flowing to the low-temperature contact~\cite{Levitov2004}, namely
\begin{equation}\label{eq:excitation-flow}
    j = \sum_{\tau=\uparrow,\downarrow} \left(j_{\text{e}\tau}+j_{\text{h}\tau}\right) = \sum_{i=\text{e,h}}\left(j_{i\uparrow}+j_{i\downarrow}\right).
\end{equation}
Importantly, this total flow of excitations includes all particles (electrons and holes), irrespective of their charge and spin.
The opacity of the conductor makes transport happen via uncorrelated single-particle tunneling events, which means that, in this limit, the current fluctuations simply count how many excitations per unit time, electrons or holes, travel from the hot to the cold contact~\cite{Reydellet2003Apr,Ol'khovskaya2008Oct,Vanevic2012Dec,Battista2014Aug}.
The factor $\kB T_\L\ln2$, which was associated to the degeneracy of the transported excitations in Ref.~\cite{Larocque2020}, stems from the influxes $j_{i\tau}$, all of which take the same value for a temperature bias because the electron-hole or spin symmetry is not broken.
This factor takes the role of an effective noise temperature when recognizing Eq.~\eqref{eq:delta-T-noise} as a generalized fluctuation-dissipation theorem~\cite{Sukhorukov1999,Larocque2020}.

The spin-current fluctuations are essentially identical to the charge current  fluctuations in Eq.~\eqref{eq:delta-T-noise}, the difference being the quantity transported. This leads to a different prefactor, namely
\begin{equation}\label{eq:delta-T-spin-noise}
    S^\Sigma (0) \approx \frac{8}{h}\left(\frac\hbar2\right)^2Dk_\text{B}T_\text{L}\ln 2 = 2\left(\frac\hbar2\right)^2 j.
\end{equation}
We highlight that the flow of excitations $j$ accounts for both spin-$\uparrow$ and spin-$\downarrow$ excitations flowing from left to right, irrespective of their spin.

\subsection{Charge current noise due to spin bias}\label{subsec:charge-noise-spin-bias}
Instead of generating the nonequilibrium noise with a temperature bias, here we set $T_\L=T_\R\to0$ and choose a spin bias. To this end, we consider a spin-nondegenerate setup in which the two spin populations in each reservoir $\alpha$ have a finite energy separation $\sigma_{\alpha}$ between spin-$\uparrow$ and spin-$\downarrow$ electrons.
Such setups can be realized by injecting spin currents in a normal metal using ferromagnetic contacting~\cite{Takahashi2008Mar}.
The occupation probability of electrons injected into the conductor is then spin-dependent and is given by Eqs.~\eqref{eq:Fermi_functions} and~\eqref{eq:chemical_spin}.
The energy separation $\sigma_\alpha$ acts as a spin-dependent chemical potential and allows the driving of spin currents through the conductor. 
The zero-charge-current condition can for instance be achieved as illustrated in Fig.~\subfigref{fig:D-uniform}{b}, with $\Delta\sigma$ finite and $\sigma_\R = 0$. Then, the flow of spin-$\uparrow$ excitations above $\mu_\L$ (electrons) is perfectly balanced by the flow of spin-$\downarrow$ excitations below $\mu_\L$ (holes). Hence, the average charge current is zero $I=-e\sum_{\tau=\uparrow,\downarrow} \left(j_{\text{e}\tau}-j_{\text{h}\tau}\right)=0$, whereas the spin current is finite.
By using Eq.~\eqref{eq:shot-general}, we then find that the charge noise in the absence of charge current becomes
\begin{align}\label{eq:delta-spin-noise}
    S^{I}(0) \equiv S^{I}_\text{sh}(0) & = 2\frac{e^2}{h}D(1-D)\left|\Delta\sigma\right|\nonumber\\
    &= 2e^2 (1-D) j,
\end{align}
very similar to the expressions of Eqs.~\eqref{eq:delta-T-noise} and \eqref{eq:delta-T-spin-noise}. Here, $j= D|\Delta\sigma|/h$, as defined in Eq.~\eqref{eq:excitation-flow}, is the total flow of excitations from left to right. 
A similar situation for zero-current charge noise driven by spin precession has recently been analyzed, the common feature being the breaking of spin degeneracy~\cite{Ludwig2020}.
At this point, we note that, once one has access to the transmission $D$, the zero-current noise can be used as an additional tool to probe the spin imbalance between the reservoirs \cite{Meair2011Aug, Arakawa2015Jan}. 

\subsection{Spin current noise due to voltage bias}\label{subsec:spin-noise-charge-bias}
We now consider the zero-current fluctuations of the spin current.
The conditions for zero spin current can be achieved as illustrated in Fig.~\subfigref{fig:D-uniform}{c}, where $T_\L=T_\R\to 0$ and there are no spin imbalances $\sigma_\alpha=0$.
With a potential bias $\Delta\mu$ between the contacts, the average spin current vanishes due to spin degeneracy: The spin-$\uparrow$ electrons flowing from left to right are perfectly balanced by the spin-$\downarrow$ electrons flowing in the same direction, namely $\Sigma = \frac{\hbar}2\sum_{i=\text{e,h}} \left(j_{i\uparrow}-j_{i\downarrow}\right)=0$, 
while there is a finite charge current from left to right. 
Note that the same situation would hold with a nonvanishing but equal spin imbalance present in both contacts, $\sigma_\text{L}=\sigma_\text{R}\neq0$.

Still, as the two spin channels are completely independent, the compensating spin-$\uparrow$ and spin-$\downarrow$ flows are partitioned by the conductor's finite transmission, leading to the spin current shot noise
\begin{align}\label{eq:delta-mu-noise}
    S^{\Sigma}(0) \equiv S^{\Sigma}_\text{sh}(0) & =\frac{\hbar}{2\pi}D(1-D)\left|\Delta\mu\right|\nonumber\\
    & =2\left(\frac\hbar2\right)^2(1-D)j,
\end{align}
which, as in Eqs.~\eqref{eq:delta-T-noise}, \eqref{eq:delta-T-spin-noise} and \eqref{eq:delta-spin-noise}, is again proportional to the bias applied to the system, in this case $\Delta\mu$.
Once more, the spin fluctuations are proportional to the total flow of excitations $j=D|\Delta\mu|/h$ being transferred from left to right~\cite{Martin1992Jan}.

\subsection{Heat noise}\label{subsec:heat-noise}
Unlike the charge and spin currents, it is not possible to reach the zero-heat-current condition by applying only one bias while using a uniform transmission.
In particular, there must be a temperature bias between the contacts.
The additional bias is required to achieve the zero-heat-current condition.
When a voltage bias is used to make the heat current in the left contact $J_\L$ vanish, the required chemical potential difference reads
\begin{equation}
 \Delta\mu_J\equiv\Delta\mu|_{J_\text{L}=0}= \pm \frac{\pi}{\sqrt{3}} k_\text{B} T_\text{L}.
\end{equation}
This choice is depicted in Fig.~\subfigref{fig:D-uniform}{d}. It entails a cancellation of heat conduction $D\pi^2k_\text{B}^2(T_\text{L}^2-T_\text{R}^2)/6h$ and Joule heating $D\Delta\mu^2/2h$, or---microscopically---a cancellation of influxes of positive excess energy, $j_\text{heat}^E$, with influxes of negative excess energy, $j_\text{cool}^E$. For the constant transmission considered here, and $\mu_\L>\mu_\R$, they are given by $j_\text{cool}^E=D/h\int_{-\infty}^{\mu_\text{R}} (E-\mu_\text{L})[1-f_\text{L}(E)]dE+D/h\int_{\mu_\text{L}}^{\infty} (E-\mu_\text{L})f_\text{L}(E)dE$ and $j_\text{heat}^E=D/h\int_{\mu_\text{R}}^{\mu_\text{L}} (E-\mu_\text{L})f_\text{L}(E)dE$. We discuss this separation of negative and positive excess energy fluxes in more detail in Appendix~\ref{sec:app:zero-freq-unbound}.

The resulting heat noise, consisting of a thermal and a shot noise contribution, equals~\cite{Eriksson2021Sep}
\begin{equation}\label{eq:heat-noise-D-uniform}
    S^J(0) = \frac{4\pi^2}{3h} D(k_\text{B}T_\text{L})^3  \left[1+\frac3{\pi^2}(1-D)A\left(\frac{\pi}{\sqrt{3}}\right)\right],
\end{equation}
where $A(x)=2x^2\ln(1+e^x)-(\pi^2+x^3)/3+4x\Li_2(-e^x)-4\Li_3(-e^x)$, and $\Li_n$ is the polylogarithmic function, with $3A(\pi/\sqrt{3})/\pi^2\approx 0.45$.
Note that the heat noise cannot be expressed in terms of the same $j^E$ as the heat current, and hence the analogy to charge and spin noises cannot be established in this respect. Due to the fact that the transported excess energy enters quadratically into the noise expressions, simple relations between heat currents and noise can only be found in the classical limit, see e.g. Ref.~\cite{Agarwalla2012May}.
Note however, that also the heat noise \eqref{eq:heat-noise-D-uniform} can be written in terms of independent electron- and hole contributions due to the elastic nature of the scattering process. For the shot noise contribution, this separation into electron- and hole contributions even continues to hold in the case of ac driving~\cite{Battista2014Aug}.

The case in which a spin imbalance is used instead of the voltage bias is very similar. Indeed, the zero-heat-current condition is achieved when the left spin imbalance satisfies $\sigma_\L=2\Delta\mu_J$, while the right contact is spin-degenerate, namely $\sigma_\R=0$.
In this case, each spin population of the system satisfies the zero-heat-current condition, $J_{\L\tau}=0$.
Therefore, the total heat noise coincides with Eq.~\eqref{eq:heat-noise-D-uniform} even when a spin imbalance, rather than a voltage bias, is used to satisfy the zero-current condition.

\section{Bounds on shot noise}\label{sec:shot-thermal-noise}
The results of Sec.~\ref{sec:constant-transmission} show how the shot noise of various transport observables in the absence of average currents can arise and that it can be made arbitrarily large by increasing the bias at the origin of the nonequilibrium situation. However, as soon as one of the biases is a temperature difference, additional thermal noise necessarily arises. An important question is hence how large the nonequilibrium noise $S^X_{\text{sh}}(0)$ can be, in comparison to the thermal noise $S^X_{\text{th}}(0)$.
Here, we address this issue for a general situation going beyond the limitations of Sec.~\ref{sec:constant-transmission}: the constraint $T_\R\to 0$ is lifted, and we consider conductors where $D(E)$ has an arbitrary energy dependence and where zero current at nonequilibrium is obtained in response to different biases, one of which is a temperature bias. 

For spin-degenerate systems, a first answer was provided in Ref.~\cite{Eriksson2021Sep}.
There, some of us showed that, under the condition~\eqref{eq:zero_current_condition},
the zero-frequency heat shot noise, $S_\text{sh}^J(0)$, is generally unbounded compared to the heat thermal noise $S_\text{th}^J(0)$. 
This happens when the transmission has specific features, most importantly an energy gap.
In Ref.~\cite{Eriksson2021Sep} an \emph{ad hoc} transmission with this feature was presented, albeit with no direct physical counterpart. In Section~\ref{subsec:unbounded-heat-shot-noise}, we present experimentally relevant examples in which the heat shot noise is shown to become arbitrarily large compared to the heat thermal noise.

In stark contrast to the heat noise, it was found in Ref.~\cite{Eriksson2021Sep} that the zero-frequency charge shot noise is bounded by the thermal noise
\begin{align}\label{eq:Charge_shot_0}
   \Ssh(0) \le \Sth(0),
\end{align}
for any transmission $D(E)$ and reservoirs' temperatures $T_\alpha$.
In Section~\ref{subsec:improved-bound}, we present a tighter bound than~\eqref{eq:Charge_shot_0}, which we also extend to the spin shot noise $S_\text{sh}^\Sigma(0)$, when the zero-current condition is fulfilled by a combination of $\Delta T$ and $\Delta\sigma$.

A general constraint like Eq.~\eqref{eq:Charge_shot_0} does not exist for the situations presented in Sec.~\ref{subsec:charge-noise-spin-bias} and \ref{subsec:spin-noise-charge-bias}, where the breaking of spin degeneracy in charge transport (respectively the breaking of e-h symmetry for spin transport) yields an effectively multi-terminal conductor, where a bound is generally not expected to hold.

\subsection{Achieving unbounded heat shot noise}\label{subsec:unbounded-heat-shot-noise}

\begin{figure}[tb]
    \centering
    \includegraphics{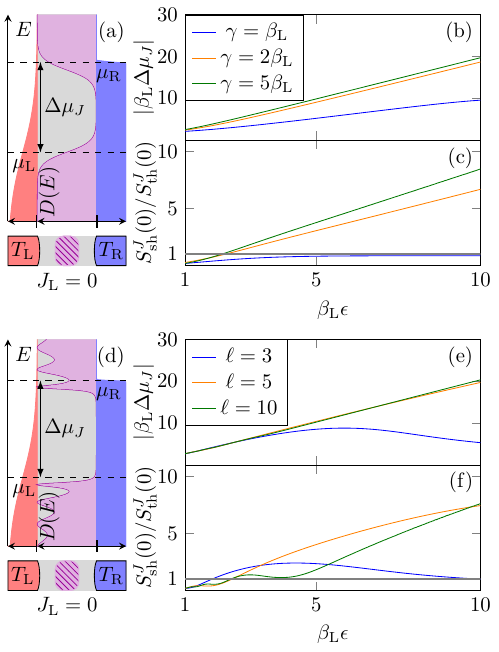}
    \caption{(a) Zero-heat-current condition for the transmission $D(E)$ from Eq.~\eqref{eq:Well-transmission} (depicted in purple). (b) Stopping voltage and (c) ratio of the zero-frequency noise components for the transmission in (a). (d) Zero-heat-current condition for the transmission $D(E)$ from Eq.~\eqref{eq:QSH_Transmission} (depicted in purple). (e) Stopping voltage and (f) ratio of the zero-frequency noise components for the transmission in (d). The heat shot noise can be much larger than the thermal noise.}
    \label{fig:zero-freq-heat-noise}
\end{figure}
Here, we demonstrate that the energy-dependent transmission of experimentally relevant conductors can lead  to zero-frequency heat shot noise that is arbitrarily large compared to the heat thermal noise, in the absence of an average heat current.
Such a behavior was observed in Ref.~\cite{Eriksson2021Sep} as a consequence of a transmission function featuring an energy gap (see in particular the Supplemental Material of Ref.~\cite{Eriksson2021Sep} for further details).

Transmissions with this type of properties are found, e.g., in conductors with helical edge states~\cite{Kane2005Nov,Konig2007Nov,Wu2006Mar}, where backscattering is induced due to etched constrictions~\cite{Sternativo2014Sep,Strunz2020} or magnetic impurities~\cite{Gresta2019Oct}.  Previously, it was highlighted in Refs.~\cite{Gresta2019Oct,Hajiloo2020Oct} that such conductors are indeed highly relevant for efficient thermoelectric applications. Moreover, we note that transmission functions complementary to the gapped ones, namely boxcar transmissions, were recently shown to produce finite current but zero shot noise scenarios~\cite{Gerry2022Apr}.

Let us first consider a ``well-shaped'' transmission probability, which can be realized in smooth or disordered helical junctions~\cite{Sternativo2014Sep}.
For the sake of concreteness, we consider the simple transmission function
\begin{equation}\label{eq:Well-transmission}
D(E-\mu_\L) = \frac1{1+e^{\gamma(E-\mu_\L)}} + \frac1{1+e^{-\gamma(E-\mu_\L-2\epsilon)}},
\end{equation}
which is illustrated in Fig.~\subfigref{fig:zero-freq-heat-noise}{a}. This transmission exhibits an energy gap of width $2\epsilon$ and the sharpness of the gap edges is determined by $\gamma$. Importantly, $\epsilon$ and $\gamma$ are two independent parameters \footnote{{Note that these parameters (and those entering other $D(E)$ as well) may depend on voltage and temperature biases in a gauge-invariant manner, see, e.g., Ref.~\cite{Benenti2017Jun}. We do not include this dependence explicitly here, since one can always tune  nearby gates such that the transmission parameters take on the desired values.}}. We next show that $\gamma$ and $\epsilon$ can be used to generate zero-current heat shot noise that can be much larger than its thermal counterpart.

First, we note that the energy gap separates the transport window into two distinct regions. Second, the sharpness of the gap edge allows a large stopping voltage $|\Delta\mu_J|$ (i.e., the voltage required to cause the heat current to vanish), which increases for an increasing gap. In turn, such a $|\Delta\mu_J|$ permits both energy regions to contribute to the heat transport, as depicted in Figs.~\subfigref{fig:zero-freq-heat-noise}{a,b}.
Furthermore, the gap edges generate finite shot noise contributions because the transmission is different from both zero and one in these regions. When a sharp edge and large $|\Delta\mu_J|$ are combined, the shot noise increases for large $\epsilon$, and approaches the linear relation $S^J_\text{sh}\propto \epsilon$, as illustrated in Fig.~\subfigref{fig:zero-freq-heat-noise}{c}. In fact, this result does not depend on the specific shape of the transmission as long as it is gapped and the gap edges are sufficiently sharp (see Appendix~\ref{sec:app:zero-freq-unbound} for details).
Therefore, the heat shot noise can be arbitrarily large compared to its thermal counterpart even under the zero-current condition.
If the gap edges are instead smooth, ($\gamma k_\text{B}T_\L\lesssim 1$), the zero-heat-current condition can be met by using only the low energy transmission region, in a similar fashion to Sec.~\ref{subsec:heat-noise}. In this case, the shot noise does not overcome the thermal noise, as illustrated by the blue curve in Fig.~\subfigref{fig:zero-freq-heat-noise}{c}.

Similar to a disordered conductor, we show that the heat shot noise can be unbounded also for a single magnetic impurity in the helical conductor. In this case, the two-terminal transmission probability displays Fabry-P\'erot-like resonances in addition to the well structure and is given by~\cite{Gresta2019Oct}
\begin{align}\label{eq:QSH_Transmission}
    &D(E-\mu_\L) = \left[1+ \frac{\sin^2(\ell r(E-\mu_\L) )}{r^2(E-\mu_\L)}\right]^{-1},\\
    &r(E) = \sqrt{\left(\frac{E-\epsilon}{\epsilon}\right)^2-1},\notag
\end{align}
where $\ell$ is a dimensionless parameter describing the effective length of the sharp impurity region and $\epsilon$ is a characteristic energy scale. Just as the transmission in Eq.~\eqref{eq:Well-transmission}, this transmission has an energy gap of $2\epsilon$, as illustrated in Fig.~\subfigref{fig:zero-freq-heat-noise}{d}. Here, however, the sharpness of the gap edges is determined by both $\epsilon$ and $\ell$.
These parameters permit a large stopping voltage $|\Delta\mu_J|$ and, consequently, a large heat shot noise, as shown in Figs.~\subfigref{fig:zero-freq-heat-noise}{e} and \subfigref{fig:zero-freq-heat-noise}{f}, respectively. In contrast to the well-like transmission setup, both stopping voltage and heat shot noise here eventually decrease at large $\epsilon$ when $\ell$ is small. This occurs because for small $\ell$, the sharpness of the gap edges decreases and only the lower transmission region becomes involved in the transport.
In addition, the Fabry-P\'erot-like resonances of the transmission produce oscillations in the noise ratio at lower $\epsilon$ and larger $\ell$. In these parameter regimes, the first few peaks of the higher energy region are involved in the transport due to the stopping voltage, as illustrated in Fig.~\subfigref{fig:zero-freq-heat-noise}{d}. Instead, at larger $\epsilon$ or lower $\ell$ these oscillations disappear because only the gap edge participates in the transport.

We conclude this section by noticing that the separation of transport into two separate energy windows can also be achieved using superconducting contacts. Indeed, the quasiparticles responsible for heat transport cannot be transmitted in the gapped energy region of the superconductor. Therefore, we expect that large heat shot noise \emph{exceeding the heat thermal noise} in the absence of heat current can be observed in superconducting devices~\cite{Marchegiani2020}.

\subsection{Bounds on zero-frequency charge and spin noise}\label{subsec:improved-bound}
In contrast to the fluctuations discussed in Secs.~\ref{subsec:charge-noise-spin-bias}, \ref{subsec:spin-noise-charge-bias}, and \ref{subsec:unbounded-heat-shot-noise}, the charge fluctuations without spin bias and the spin fluctuations without voltage bias satisfy a bound, which was already mentioned in Eq.~\eqref{eq:Charge_shot_0} for the charge noise components. Here, we improve this inequality by finding a stricter upper bound and we extend it to the spin fluctuations.
To do this, we denote
\begin{equation}\label{eq:thermal-noise-contribution}
    \cT_\alpha^X = \sum_\tau\!\int\! dE \frac{2x^2}{h} D(E) \fat[1-\fat]
\end{equation}
the contribution of reservoir $\alpha$ to the zero-frequency thermal noise, i.e.,  $S_\text{th}^X(0)=\cT^X_{\L}+\cT^X_{\R}$.
In spin-degenerate systems, in which the nonequilibrium condition is determined only by a temperature and a voltage bias (and no spin splitting is present), we consider the charge noise in the absence of an average charge current.
We find (see Appendix~\ref{sec:app:zero-freq-bound} for the derivation) that the shot noise is bounded as
\begin{equation}
    \Ssh(0) \le \cT^I_\text{h} - \cT^I_\text{c}, \label{eq:bound-shot-charge}
\end{equation}
where $\cT^I_\text{h}$ and $\cT^I_\text{c}$ are the thermal noise contributions of the \emph{hot} and \emph{cold} reservoirs respectively, e.g., $\text{h} = \L$ and $\text{c}=\R$ if $T_\R < T_\L$.

The bound~\eqref{eq:bound-shot-charge} is tighter than that in Eq.~\eqref{eq:Charge_shot_0} from Ref.~\cite{Eriksson2021Sep}. This is so since $S_\text{th}^X(0) = \cT^X_\L + \cT^X_\R$ and $\cT^X_\alpha \ge 0$. Equality in Eq.~\eqref{eq:bound-shot-charge} is approached when the maximum value of the transmission satisfies $D_\text{max}\ll 1$ and, at the same time, $D(E)$ is finite in an energy interval $\delta$ much smaller than the hot temperature, e.g., $\delta \ll \kB T_{\L}$ for $T_{\L}>T_{\R}$.
In the case of equal thermal noise contributions, $\cT^I_\L = \cT^I_\R$, the shot noise vanishes under the zero-current condition~\eqref{eq:zero_current_condition}.
Indeed, even though the same thermal fluctuations can be achieved at different temperatures, satisfying both $\cT^I_\L = \cT^I_\R$ and zero charge current is only possible in equilibrium. Then, the total noise simply reduces to equilibrium thermal noise.

\begin{figure}[tb]
    \centering
    \includegraphics{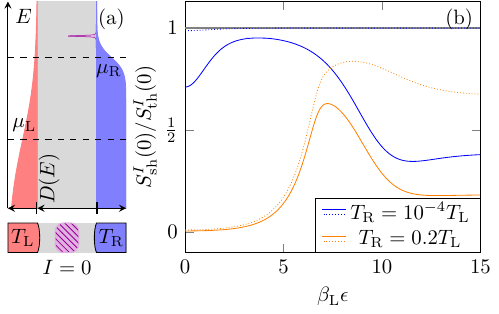}
    \caption{(a) Zero-charge-current condition for the Lorentzian transmission $D(E)$ from Eq.~\eqref{eq:lorentzian} (depicted in purple), with fixed $\beta_\L{\Gamma}=10^{-2}$ and $D_0=0.5$.  (b) Charge noise ratio $S_\text{sh}^I(0)/S_\text{th}^I(0)$ for the Lorentzian transmission as a function of the resonance peak position $\epsilon$ and for two different temperatures of the cold reservoir. Dotted lines correspond to the bound in Eq.~\eqref{eq:bound-shot-charge}. The solid, grey line corresponds to the bound in Eq.~\eqref{eq:Charge_shot_0}.}
    \label{fig:new-bound}
\end{figure}

We illustrate the improved bound~\eqref{eq:bound-shot-charge} in Fig.~\ref{fig:new-bound}, where we plot the noise ratio $S_\text{sh}^I(0)/S_\text{th}^I(0)$ obtained with the Lorentzian transmission function
\begin{equation}
    D(E-\mu_\L)=D_0\frac{{\Gamma}^2}{{\Gamma}^2+(E-\mu_\L-\epsilon)^2}\,.
    \label{eq:lorentzian}
\end{equation}
As shown in Ref.~\cite{Eriksson2021Sep}, this transmission is particularly useful as it interpolates between two different regimes, from a sharp resonance to an almost constant transmission. Fig.~\subfigref{fig:new-bound}{b} is obtained by selecting a sharp resonance with $\beta_\L{\Gamma}=10^{-2}$, so that it is possible to get closer to the bound (compared to a case with a larger width).
As expected, when the temperature of the cold reservoir $T_\R$ is negligible, the bound~\eqref{eq:bound-shot-charge} reduces to~\eqref{eq:Charge_shot_0}, as $\Theta^I_\text{R}\approx 0$ in this regime. In contrast, when $T_\R$ is sizable, we observe that the improved bound~\eqref{eq:bound-shot-charge} captures much better the behavior of the shot noise, whereas Eq.~\eqref{eq:Charge_shot_0} clearly fails to do so. The large differences between solid and dotted lines observed at large $\epsilon$ are explained by realizing that, in this regime, the resonance peak lies outside the bias window and the tails of the Lorentzian behave as a constant transmission, for which the bound cannot be approached.

Similarly to the charge noise, we next consider the spin-current fluctuations in spin-nondegenerate systems in the absence of voltage bias. In this case, the nonequilibrium condition is determined by a combination of the temperature bias and the spin imbalances of the reservoirs.
Again, when the average spin current vanishes, we find that the corresponding shot noise is bounded as
\begin{equation}\label{eq:bound-shot-spin}
    S^\Sigma_\text{sh}(0) \leq \Theta^\Sigma_\text{h}-\Theta^\Sigma_\text{c}\le S_\text{th}^\Sigma(0)=\cT^\Sigma_\text{h}+\cT^\Sigma_\text{c}.
\end{equation}

We stress here that the bounds in Eqs.~\eqref{eq:bound-shot-charge} and ~\eqref{eq:bound-shot-spin} hold for \emph{any} transmission function $D(E)$ and \emph{any} temperatures $T_\L,T_\R$, as long as the zero-current condition is fulfilled.
These inequalities are due to both charge and spin being energy-independent electronic properties as well as the induced symmetries in the different setups. More specifically, in contrast to heat, the transported charge and spin do not depend on the energy at which the particle tunnels through the conductor. Furthermore, for the charge fluctuations, the system is spin-invariant, namely $f_{\alpha\uparrow}(E) = f_{\alpha\downarrow}(E)$, while for the spin fluctuations, the system is electron-hole symmetric, i.e., $f_{\alpha\uparrow}(E) = 1-f_{\alpha\downarrow}(-E)$.
It is therefore interesting to note that violations of the bound~\eqref{eq:bound-shot-charge} could be a signature of the conductor breaking the spin degeneracy.
Note that breaking the above symmetries makes the system equivalent to a multi-terminal device in which there is one terminal for each index pair $\alpha\tau$. In this case, the additional control parameters provided by the lack of the symmetry constraints can be combined to break the bounds \eqref{eq:bound-shot-charge} and~\eqref{eq:bound-shot-spin}.

The fact that both charge and spin are energy-independent quantities makes the charge and spin fluctuations proportional to each other, see Eq.~\eqref{eq:noise-general}. This implies that, when the charge shot noise is bounded according to~\eqref{eq:bound-shot-charge}, the spin shot noise is also bounded according to~\eqref{eq:bound-shot-spin}, under the same nonequilibrium conditions  (which also lead to zero spin current, $\Sigma=0$). Vice versa, when~\eqref{eq:bound-shot-spin} is satisfied, the charge shot noise is bounded, according to~\eqref{eq:bound-shot-charge}, even though the nonequilibrium conditions leading to $\Sigma=0$ do not imply $I=0$.


\section{Finite-frequency noise in the absence of currents}\label{sec:finite-frequency}

We have demonstrated in Sec.~\ref{sec:shot-thermal-noise} that, at zero frequency, the thermal noise, necessarily arising when the conductor is biased by a temperature difference, sets a bound on the shot noise in the absence of an average charge or spin current. This holds as long as the transported quantity is energy-independent and the conductor is truly two-terminal. In contrast, this is not the case for heat transport or when the two-terminal conductor (effectively) acts as a multi-terminal conductor. A natural next question is therefore whether the $\omega=0$ bound for the charge shot noise~\eqref{eq:bound-shot-charge} exists even for finite-frequency noise $\omega \neq 0$.

Finite-frequency noise in mesoscopic systems has long been studied, in particular in connection with the so-called quantum-noise regime~\cite{Nazarov2003NATO}. It can be linked to the transition rates of a two-level system or, more generally, to the emission and absorption spectra of a conductor coupled to an electromagnetic environment. In the latter case, the finite-frequency noise is expressed as a weighted sum over single-particle transitions associated with the emission or absorption of a photon of energy $\hbar\omega$~\cite{Gavish2001}. Simultaneously to our work, a finite-frequency generalization of delta-$T$ noise~\cite{Hubler2022Oct} was investigated, which shows the interest in finite-frequency noise under zero-current conditions.

To investigate the finite frequency and zero current noise, we separate in this section the finite-frequency nonequilibrium partitioning noise and the remaining terms according to their physical characteristics, in the same spirit as in Eq.~\eqref{eq:noise-general}.
Then, we demonstrate that the zero-current shot noise is indeed limited by the remaining terms also at finite frequency, see Eq.~\eqref{eq:finite-freq-bound}.
Notably, the zero-frequency limit of this bound is tighter than Eq.~\eqref{eq:Charge_shot_0}, but looser than Eq.~\eqref{eq:bound-shot-charge}.
We thereby corroborate the generality of the bound on charge shot noise in the absence of current and identify detailed requirements for the bound to hold.

\subsection{Scattering approach to finite-frequency noise}\label{subsec:scattering-approach-finite-frequency}

Here, we present analytical results for the autocorrelation charge shot noise in contact $\alpha=\text{L}$ at finite frequency, $\omega$. In the limit of $\omega\to 0$, this yields the expressions in Eq.~\eqref{eq:noise-general}, which 
were the basis of the analyses of Secs.~\ref{sec:constant-transmission} and \ref{sec:shot-thermal-noise}. 
At finite frequency, the total noise can be decomposed into an absorption [$S^{X,-}(\omega)$] and an emission [$S^{X,+}(\omega)$] contribution, which are associated with the absorption and emission of an energy quantum $\hbar\omega$, respectively.
The symmetrized noise in Eq.~\eqref{eq:Noise_def} is simply the sum of absorption and emission contributions, namely $S^X(\omega)= S^{X,-}(\omega)+S^{X,+}(\omega)$~{\cite{BlanterButtikerPhysRep00,Moskalets2011Sep}}.
However, it is generally possible to consider and measure nonsymmetric combinations of absorption and emission noise~\cite{Lesovik1997Feb, Glattli2009Jun}.

To be able to compare our results to the previous sections, we divide both contributions into two expressions: the shot-noise expression of interest, yielding Eq.~\eqref{eq:shot-general} in the limit $\omega\to 0$ and the remaining terms that are not directly related to a nonequilibrium partitioning noise. This division is not unique, cf. for instance Ref.~\cite{Hubler2022Oct}. In our case, we define   the shot noise contribution of the absorption noise as
\begin{align}
       \Ssh[I,-](\omega)=& \frac{2e^2}{h}\!\int\! dE  D(E)[1-D(E+\hbar\omega)][\fL-\fR]\nonumber\\
       &\times [\FL - \FR].
    \label{eq:shot-finite-freq}
\end{align}
and the emission shot noise is given by $\Ssh[I,+](\omega)=\Ssh[I,-](-\omega)$. The symmetric form of Eq.~\eqref{eq:shot-finite-freq} with respect to the exchange $\R\leftrightarrow\L$ implies that the shot noise component does not depend on the reservoir in which the finite-frequency noise is measured, consistent with the role of the conductor as the source of partitioning currents coming from the two reservoirs.

The remaining terms, which become $\Sth(0)$ in the limit $\omega \to 0$, are given by $\Sr[I](\omega)=\Sr[I,-](\omega)+\Sr[I,+](\omega)$ with
\begin{subequations}
\label{eq:finite_freq_full}
    \begin{align}
\Sr[I,-](\omega)  = & \cT^{-}_\L(\omega)+\cT^{-}_\R(\omega)+\cA^{-}(\omega)+\cR^{-}(\omega),\\
\cT^{-}_\alpha(\omega)=  &  \frac{2e^2}{h}  \!\int\!dE \left[\frac{D(E)+D(E+\hbar\omega)}{2}\right]\nonumber\\
    &\times\{\fa[1-\Fal]\}\label{eq:finite-freq-TL-absorption},\\
\cR^-(\omega)  = & \frac{2e^2}{h}\!\int\! dE \left|s_\text{LL}(E)-s_\text{LL}(E+\hbar\omega)\right|^2\nonumber\\
    &\times \{\fL[1-\FL]\}, \label{eq:thermal-R-component-absorption}\\
\cA^-(\omega)  = & \frac{2e^2}{h}  \!\int\!dE \left[\frac{D(E)-D(E+\hbar\omega)}2\right]\nonumber\\
         &\times \{\fR[1-\FL]\nonumber\\
    & -\fL[1-\FR]\},\label{eq:thermal-A-component-absorption}
\end{align}
\end{subequations}
together with the emission contribution, $\Sr[I,+](\omega)=\Sr[I,-](-\omega)$.
Here, $\cT^{-}_\alpha(\omega)$ is the contribution that yields the thermal noise, stemming from the thermal excitations in contact $\alpha$.  In contrast, the terms $\cR^{-}(\omega)$ and $\cA^{-}(\omega)$ vanish at $\omega = 0$ or when the transmission function is energy independent, $D(E)\equiv D$. The contribution $\cR^-(\omega)$ contains the factor $f_\L(1-f_\L)$, typical of thermal fluctuations, and does thus not vanish in equilibrium. Moreover, $s_\text{LL}(E)$ is the scattering matrix element describing the reflection amplitude of reservoir $\L$, see Eq.~\eqref{eq:smatrix}.  
By contrast, the term $\cA^-(\omega)$ vanishes in equilibrium, but does not contain the partitioning factor $D(1-D)$. It does therefore not have the characteristic partition property of the shot noise. The system considered here is spin-degenerate, which produces the factors of $2$ in the above noise expressions.

Note that at zero frequency, the noise obeys conservation laws which relate the autocorrelation noise with the cross-correlation noise. These laws imply that in the two-terminal configuration, the autocorrelations in Eq.~\eqref{eq:noise-general} are not independent of the cross-correlations. In contrast, at finite frequency, such conservation laws are absent and finite frequency auto- and cross-correlation noises can have different features~\footnote{In addition, note that at high enough frequency, contributions arising from displacement currents have to be taken into account too~\cite{BlanterButtikerPhysRep00}.}. Here, we restrict our study to the properties of the charge current autocorrelator under the zero-current condition, corresponding to the finite-frequency delta-$T$ noise observed by measuring the charge noise in a single reservoir. 


\subsection{Bound for finite-frequency charge noise}
In this section, we present a bound for the finite-frequency shot noise which extends the zero-frequency shot noise bound in Sec.~\ref{subsec:scattering-approach-finite-frequency} (see Appendix \ref{sec:app:finite-freq-bound} for its detailed  derivation). Interestingly, the absorption part of the zero-current charge shot noise measured from the autocorrelation in the left contact obeys a bound, which we find to be
\begin{equation}\label{eq:finite-frequency-absorption-bound}
    S_\text{sh}^{I, -}(\omega) \leq  \cA^-(\omega) + {\cT}^-_\R(\omega).
\end{equation}
This bound holds as long as the noise is measured in the colder reservoir, e.g., the left reservoir for $T_\L<T_\R$.
In contrast, an equivalent expression for a bound does \textit{not} hold for the emission noise $S^{I,+}_\text{sh}(\omega)$, because the thermal contributions $\cT^+_{\alpha}(\omega)$ are much smaller than their absorption counterparts. Indeed, $\cT^+_{\alpha}(\omega)$ contains the product $f_\alpha(E+\hbar\omega)[1-f_\alpha(E)]$, which decreases exponentially in $\omega$. This leads to a smaller $S^{I,+}_\text{rest}(\omega)$, and allows the shot noise to be larger than the remaining noise.

However, for the total finite-frequency shot noise at zero current, including both the absorption and emission part, we find an equivalent bound
\begin{align}
	S^I_\text{sh}(\omega) \leq  \cA(\omega) + {\cT}_\R(\omega),\label{eq:finite-freq-bound}
\end{align}
where the sum of contributions $\cA(\omega)=\cA^{-}(\omega)+\cA^{+}(\omega)$ and ${\cT}_\R(\omega)={\cT}^{-}_\R(\omega)+{\cT}^{+}_\R(\omega)$ appear. This bound holds again provided that the noise is measured in the colder reservoir, here taken as $\L$, i.e. for $T_\L<T_\R$. 

The bound~\eqref{eq:finite-freq-bound} implies that, in the absence of charge current, the finite-frequency charge shot noise is always smaller than the remaining noise contributions, $\Ssh[I](\omega)\leq\Sr[I](\omega)$, when the noise measurement is performed in the colder reservoir. This can be understood from the fact that $\cR(\omega)$ as well as $\cT_\text{L}(\omega)$ are always positive quantities.
Taking the limit $\omega\to 0$ in Eq.~\eqref{eq:finite-freq-bound}, we obtain the bound $\Ssh[I](0) \le \cT_\text{h}$. This is a weaker zero-frequency bound compared to the one that we previously found in Eq.~\eqref{eq:bound-shot-charge}.
This happens because the frequency $\omega$ acts as an additional control parameter by broadening the distributions $f_\alpha(E)[1-f_\alpha(E+\hbar\omega)]$ entering the noise.
In particular, increasing $\omega$ makes $f_\alpha(E)[1-f_\alpha(E + \hbar\omega)]$ larger.
This feature can be used to increase the finite-frequency shot noise, as illustrated in Fig.~\subfigref{fig:finite-freq-noise-cold}{b}. As a consequence, the finite-frequency extension of the zero-frequency bound~\eqref{eq:bound-shot-charge} does not hold, whereas the weaker bound in Eq.~\eqref{eq:finite-freq-bound} does.

\subsubsection{Approaching the finite-frequency bound}

\begin{figure}[b!]
    \centering
    \includegraphics{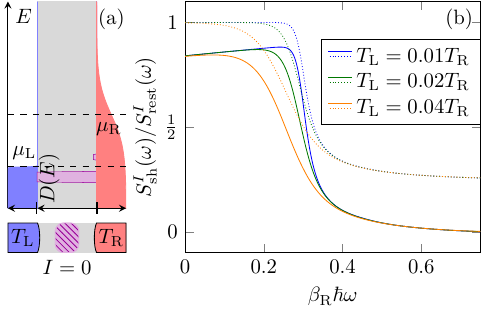}
    \caption{(a) Zero-charge-current condition for the transmission $D(E)$ in Eq.~\eqref{eq:Transmission_forFig2} (depicted in purple). (b) Ratio of the finite frequency noise components. Noise is measured in the left, cold reservoir, $T_\L<T_\R$. Dotted lines correspond to the bound~\eqref{eq:finite-freq-bound}.}
    \label{fig:finite-freq-noise-cold}
\end{figure}

With an appropriate choice of transmission function $D(E)$, it is possible to approach the bound~\eqref{eq:finite-freq-bound} at finite frequency, as shown in Fig.~\ref{fig:finite-freq-noise-cold}.
As a concrete realization of this, we consider a transmission composed of two separated energy windows: The window at lower energy has perfect transmission, $D=1$, whereas the higher energy window has a weak transmission, $D\ll1$, namely
\begin{equation}
\label{eq:Transmission_forFig2}
    D(E) = \left\{\begin{array}{ll}
        1 &    \text{if}\:E-E_0\in[0, 2\delta], \\
        0.05 & \text{if}\:E-E_1\in[0, \delta],  \\
        0 & \text{elsewhere},
    \end{array}\right.
\end{equation}
with $E_0 = \mu_\L-0.9\kB T_\R$, $E_1 = \mu_\L + 0.45\kB T_\R$ and $\delta = 0.3\kB T_\R$.
The smaller the weak transmission, the more the noise ratio $S^I_\text{sh}(\omega)/\Sr[I](\omega)$ approaches the bound. For frequencies such that $\hbar\omega<-(E_0+2\delta)$, i.e., smaller than the separation between $\mu_\L$ and the edge of the low energy window, the shot noise increases while the thermal noise essentially remains constant at low temperatures. This happens because the finite-frequency fluctuations increase the energy range of particles contributing to the noise. However, since the left reservoir is cold and the hot reservoir is subject to a high thermovoltage $\mu_\R$ [see Fig.~\subfigref{fig:finite-freq-noise-cold}{a}], the occupation number of the reservoirs in the transport region is close to either 0 or 1, leading to small thermal fluctuations.
By contrast, for $\hbar\omega>-(E_0+2\delta)$, the thermal noise increases rapidly to its maximum value, leading to a reduction of the noise ratio.
Indeed,  in this case the thermal fluctuations of the cold reservoir are large, because they can involve the occupations below and above $\mu_\L$, namely $f_\L(E)[1-f_\L(E + \hbar\omega)]$.
\subsubsection{Breaking the finite-frequency bound in the hot contact}
\begin{figure}[t!]
    \centering
    \includegraphics{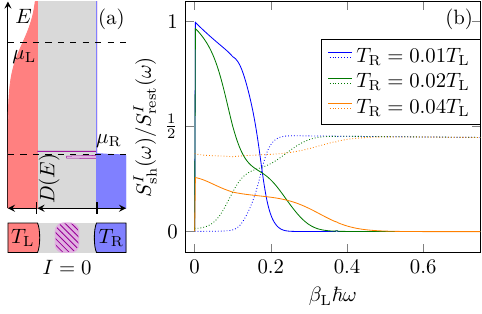}
    \caption{(a) Zero-charge-current condition for the transmission $D(E)$ from Eq.~\eqref{eq:Transmission_forFig3} (depicted in purple). (b) Ratio of the finite frequency noise components. Noise is measured in the left, hot reservoir, $T_\R<T_\L$. Dotted lines correspond to the bound~\eqref{eq:finite-freq-bound}, which is clearly broken, as expected for measurements in the hot reservoir.}
    \label{fig:finite-freq-noise-hot}
\end{figure}
An important requirement to reach the finite frequency bound in Eq.~\eqref{eq:finite-freq-bound} is that the noise is measured in the \textit{colder} reservoir, i.e., $T_\L<T_\R$.
If instead the noise is measured in the hotter reservoir $T_\L>T_\R$, the bound~\eqref{eq:finite-freq-bound} does not hold.
To show this, we consider the transmission $D(E)$ depicted in Fig.~\subfigref{fig:finite-freq-noise-hot}{a}, made out of two energy windows.
Specifically, we take
\begin{equation}
\label{eq:Transmission_forFig3}
    D(E) = \left\{\begin{array}{ll}
        1/2 & \text{if}\: E-E_0\in[0, 0.1\kB T_\L],\\
        1 &  \text{if}\: E-E_1\in[0,  \delta],\\
        0 & \text{elsewhere},
    \end{array}\right.
\end{equation}
with $E_0 = \mu_\L-6.96\kB T_\L$, $E_1 = \mu_\L-6.56\kB T_\L$ and $\delta = 5\cdot 10^{-5} \kB T_\L$.
In this setup, $S^I_\text{sh}(\omega)/\Sr[I](\omega)$ increases rapidly as the frequency becomes comparable to the width $\delta$ of the narrow window transmission, and decreases for larger $\omega$. This is illustrated in Fig.~\subfigref{fig:finite-freq-noise-hot}{b}.
Note that at zero frequency the bound \eqref{eq:bound-shot-charge} is still satisfied, and, in this particular case, limits the shot noise to being much smaller than the thermal noise.
Indeed, the shot noise is produced only in the lower transmission window, where $f_\L$ and $f_\R$ are close due to the large thermovoltage. There, the shot noise scales as $(f_\L-f_\R)^2$, whereas the thermal noise scales as $f_\L(1-f_\L)\approx|f_\L-f_\R|$.
We can understand the behavior of $S^I_\text{sh}(\omega)/\Sr[I](\omega)$ with the same arguments as in the case for the noise in the colder reservoir.
In particular, for this setup, the increase in finite frequency shot noise due to the narrow upper window allows it to overcome the bound in Eq.~\eqref{eq:finite-freq-bound}. Nonetheless, our numerical analysis does not show any instance of the charge shot noise ever being greater than the thermal noise. Therefore, we conjecture that the previously derived bound~\eqref{eq:Charge_shot_0} should hold at finite frequency for any temperature bias as well.

\section{\label{sec:measurement_scheme} Shot noise measurement scheme}
In this section, we address the possibility of experimentally verifying the bounds on charge shot noise in the absence of average charge current both in the zero and the finite-frequency regime.

To this end, we start by proposing a simple measurement scheme that allows to determine the finite-frequency charge shot noise, and to estimate the upper bounds in Eqs.~\eqref{eq:finite-frequency-absorption-bound} and~\eqref{eq:finite-freq-bound}.
The procedure involves measuring the charge current fluctuations with different choices for the electrochemical potentials and temperatures in both contacts. We therefore explicitly write the noise dependence on these quantities as $S^I(\omega; \{\mu_\text{L}, T_\text{L}\}; \{\mu_\text{R}, T_\text{R}\})$. 
Here, the first (second) curly brackets indicate the electrochemical potential and temperature of the left (right) contact.
In the following, we set $\lambda \equiv \{\mu_\text{L}, T_\text{L}\}$ and $\rho \equiv \{\mu_\text{R}, T_\text{R}\}$ for compactness.

To obtain the shot noise, four measurements are required.
The first finite-frequency noise measurement is performed under the desired out-of-equilibrium condition, yielding $S^I(\omega; \lambda; \rho)$.
Using the decompositions of Eqs.~\eqref{eq:shot-finite-freq} and \eqref{eq:finite_freq_full}, the measured noise reads
\begin{equation}\label{eq:S(lambda,rho)}
\begin{split}
    S^I(\omega; \lambda; \rho) = &S_\text{sh}^I(\omega; \lambda; \rho) + \Theta_\text{L}(\omega) + \Theta_\text{R}(\omega) \\ &+\mathcal{R}(\omega; \lambda) + \mathcal{A}(\omega; \lambda; \rho).
\end{split}
\end{equation}
Notably, the shot noise is symmetric under the exchange of the contact electrochemical potentials and temperatures, namely $S_\text{sh}^I(\omega; \lambda; \rho) = S_\text{sh}^I(\omega; \rho; \lambda)$. In contrast, the quantity $\mathcal{A}$ is antisymmetric: $\mathcal{A}(\omega; \lambda; \rho) = -\mathcal{A}(\omega; \rho; \lambda)$.
These properties suggest a second noise measurement, which is performed with exchanged electrochemical potentials and temperatures, namely by inverting both voltage and temperature biases across the device. One then finds the noise
\begin{equation}\label{eq:S(rho,lambda)}
\begin{split}
    S^I(\omega; \rho; \lambda) = &S_\text{sh}^I(\omega; \lambda; \rho) + \Theta_\text{L}(\omega) + \Theta_\text{R}(\omega) \\ &+\mathcal{R}(\omega; \rho) - \mathcal{A}(\omega; \lambda; \rho).
\end{split}
\end{equation}
Furthermore, both $S_\text{sh}^I$ and $\mathcal{A}$ vanish under equilibrium conditions (see Eqs.~\eqref{eq:shot-finite-freq} and \eqref{eq:thermal-A-component-absorption} respectively). Two final measurements are to be performed when the contacts have the same electrochemical potentials and temperatures, which gives
\begin{align}
    S^I(\omega; \lambda; \lambda) &= 2\Theta_\text{L}(\omega) +\mathcal{R}(\omega; \lambda), \label{eq:S(lambda,lambda)}\\
    S^I(\omega; \rho; \rho) &= 2\Theta_\text{R}(\omega) +\mathcal{R}(\omega; \rho) \label{eq:S(rho,rho)}.
\end{align}

The difference between the out-of-equilibrium fluctuations in Eqs.~\eqref{eq:S(lambda,rho)} and~\eqref{eq:S(rho,lambda)}, combined with the equilibrium noise in Eqs.~\eqref{eq:S(lambda,lambda)} and~\eqref{eq:S(rho,rho)} is proportional to the shot noise. More specifically, we have
\begin{equation}
\begin{split}
    S^I_\text{sh}(\omega; \lambda; \rho) = \frac12&\left[S^I(\omega; \lambda; \rho) + S^I(\omega; \rho; \lambda) \right.\\
    &\,\left.-S^I(\omega; \lambda; \lambda) - S^I(\omega; \rho; \rho)\right].
\end{split}
\end{equation}
This result means that the finite-frequency shot noise can be experimentally accessed.
The same noise measurements give also an estimate of the upper bound in Eq.~\eqref{eq:finite-freq-bound} through
\begin{equation}\label{eq:bound-estimator}
\begin{split}
    \mathcal{A}(\omega) + \Theta_\text{R}(\omega) + \frac{\mathcal{R}(\omega)}{2} = \frac12&\left[S^I(\omega; \lambda; \rho) - S^I(\omega; \rho; \lambda) \right.\\
    &\,\left.  + S^I(\omega; \rho; \rho)\right].
\end{split}
\end{equation}
Here, $\mathcal{A}(\omega)\equiv \mathcal{A}(\omega; \lambda; \rho)$ and $\mathcal{R}(\omega)\equiv\mathcal{R}(\omega; \lambda)$ are the noise components in the desired out-of-equilibrium condition.
Since $\mathcal{R}(\omega)$ is non-negative, Eq.~\eqref{eq:bound-estimator} is always greater than the zero-current charge shot noise $S_\text{sh}^I(\omega)$ [see Eq.~\eqref{eq:finite-freq-bound}], and, at the same time, smaller than the remaining noise $S_\text{rest}^I(\omega)$.

Even though we discussed the measurement scheme for the symmetrized noise, applying the same procedure to the absorption noise gives the same results.
Therefore, the four noise measurements discussed above can be used to verify the bounds of Eqs.~\eqref{eq:finite-frequency-absorption-bound} and~\eqref{eq:finite-freq-bound}.
Furthermore, at zero frequency, where $\mathcal{R}(0;\lambda)=\mathcal{A}(0; \lambda; \rho)=0$, the noise measurements in Eqs.~\eqref{eq:S(lambda,rho)}, \eqref{eq:S(lambda,lambda)}, and \eqref{eq:S(rho,rho)} allow one to determine the shot noise and each contact's contribution to the thermal noise~\cite{Hasegawa2021}. In turn, these quantities can be used to verify the bound~\eqref{eq:bound-shot-charge} of the zero-current shot noise at zero frequency.

\section{\label{sec:Summary_Conclusions} Conclusions}

We have studied noise in steady-state charge, spin, and heat transport in a two-terminal quantum conductor under a wide range of nonequilibrium conditions such that the corresponding average currents vanish. We thus extended the concept of so-called delta-$T$ noise~\cite{Sukhorukov1999,Lumbroso2018} (zero-current charge noise due to a pure temperature bias) to more generic setups of quantum transport, which are of wide interest in the fields of thermoelectrics and spintronics.

While, generally, the shot noise can be arbitrarily large, even in the absence of the corresponding average current, we have demonstrated that this is not the case for charge and spin fluctuations in spin- and electron-hole-symmetric systems, respectively.
Indeed, the shot noises are limited by the difference of their thermal counterparts for any conductor and any temperature bias.
Furthermore, we have extended this bound on the zero-current charge fluctuations to the finite-frequency regime, where it holds as long as the noise is measured in the cold contact.

We envision that the general concept of zero-current nonequilibrium noise could be developed into a versatile diagnostic tool for probing, e.g., superconducting devices, strongly correlated electron circuits~\cite{Safi2020Jul,Zhang2022} or small-scale quantum thermodynamical machines working in the absence of average heat transfer~\cite{Sanchez2013Dec,Freitas2021Mar,Sanchez2019Nov,Deghi2020Jul,Hajiloo2020OctThermo,Ciliberto2020Nov}. Of particular use is the strong dependence of the noise on temperature gradients or spin imbalances. Another extension of our work could target zero-current noise in situations where temperature fluctuations are important: Since in thermoelectric materials charge and heat transport are correlated, temperature fluctuations have been demonstrated to enhance voltage fluctuations, even in equilibrium~\cite{Tran2023Jan}.

\begin{acknowledgments}
We thank Patrick Potts for useful discussions and for pointing us to the question of finite-frequency noise, as well as Roselle Ngaloy for advice concerning spin-polarized conductors. 
We acknowledge support from the Swedish VR Vetenskapsr\r{a}det, Project No. 2018-05061 (J.S. and J.M.), the Knut and Alice Wallenberg Foundation (M.A., J.S., and L.T.), and the Nano Area of Advance at Chalmers University of Technology (C.S. and J.S.). This project has received funding from the European Union's Horizon 2020 research and innovation programme under grant agreement No 101031655 (TEAPOT). 
\end{acknowledgments}
\appendix

\section{Small temperature bias}\label{app:smallbias}
In Sec.~\ref{sec:constant-transmission} and in the Supplemental Material of Ref.~\cite{Eriksson2021Sep}, we considered the experimentally relevant~\cite{Larocque2020} situation of a large temperature bias $T_\L\gg T_\R$. On the other hand, experiments (e.g., in Ref.~\cite{Lumbroso2018}) have also been performed in the opposite regime, namely when the thermal bias $\Delta T=T_\L-T_\R>0$ is smaller than the average temperature $\bar{T}=(T_\L+\text{T}_\R)/2$. Some results for the delta-$T$ noise in this regime have appeared in previous works~\cite{Lumbroso2018,RechMartinPRL20,Schiller2022,Zhang2022}, while they have not been reported for the heat noise. 

In this Appendix, for completeness, we provide an analytical expression to leading order in $\Delta T/\bar{T}$ for the zero-current heat noise at zero frequency, $S^J(0)$ [see Eq.~\eqref{eq:noise-general}]. For convenience, we set $\mu_\L=0$ as the reference energy and we also limit ourselves to a spin-degenerate system. The heat shot noise contribution~\eqref{eq:shot-general} can be expanded as
\begin{equation}
    \label{eq:SJ_expansion}
    \Ssh[J](0) \approx\frac{4}{h}(\kB \bar{T})^3\left[S^J_0+S^J_1\left(\frac{\Delta T}{\bar{T}}\right)\right],
\end{equation}
where
\begin{equation}
    \begin{split}
        S_0^J&=F(0)\left[\frac{y}{3}\left({y}^2+\pi^2\right)\coth{\frac{y}{2}}-{y}^2-\frac{2\pi^2}{3}\right]\\
        &+\frac{k_\mathrm{B}\bar{T}}{\Gamma}F'(0)\left[y\left({y}^2+\pi^2\right)-\frac{{y}^2}{4}\left({y}^2+2\pi^2\right)\coth{\frac{y}{2}}\right]
    \end{split}
\end{equation}
and
\begin{equation}
    \begin{split}
        S_1^J&=F(0)\left[\frac{{y}^2}{2}-\frac{\pi^2}{3}y\coth{\frac{y}{2}}+\frac{{y}^2}{6}\left(\frac{{y}^2}{4}+\pi^2\right)\mathrm{csch}^2{\frac{y}{2}}\right]\\
        &+\frac{\kB \bar{T}}{\Gamma}F'(0)\left[\pi^2\left(\frac{{y}^2}{2}+\frac{7\pi^2}{15}\right)\coth{\frac{y}{2}}-\frac{y}{2}\left(3\pi^2+y^2\right)\right.\\
        &\left.\qquad\qquad\qquad-\left(\frac{3y^5}{80}+\frac{5\pi^2y^3}{24}+\frac{7\pi^4y}{30}\right)\mathrm{csch}^2{\frac{y}{2}}\right].
    \end{split}
\end{equation}
Here, $y\equiv(\mu_\L-\mu_\R)/\kB \bar{T}$ is a dimensionless bias, $F(E)\equiv D(E)[1-D(E)]$ with $F(0)=F(\mu_\L)$, and $\Gamma$ is the typical energy scale over which the transmission function varies, for instance the width of a Lorentzian. The above expressions are valid under the assumption $\kB \bar{T}\ll\Gamma$, i.e., they provide a first order correction to the constant transmission scenario, which is recovered for $F'(E)\equiv\Gamma dF/dE=0$.

Next, we find the stopping voltage $\Delta\mu_J$, i.e., the voltage needed for the heat current $J_\L$ to vanish. For a constant transmission $D(E)=D$, this voltage is obtained exactly as
\begin{equation}
    |\Delta\mu_J|=\frac{\kB \pi}{\sqrt{3}}\sqrt{T_\L^2-T_\R^2}= \frac{\kB \bar{T}\pi}{\sqrt{3}}\sqrt{\frac{2\Delta T}{\bar{T}}},
\end{equation}
and by substituting $y$ with $\Delta\mu_J/(\kB\bar{T})$ in Eq.~\eqref{eq:SJ_expansion}, we find the leading order approximation for the zero-current heat shot noise as
\begin{equation}
    \Ssh[J](0)\approx\frac{4}{h}(\kB \bar{T})^3D(1-D)\frac{\pi^2(\pi^2-6)}{27}\frac{\Delta T}{\bar{T}}\,.
\end{equation}
Moving on to a generic transmission function $D(E)$, we find instead the stopping voltage
\begin{equation}
    \Delta\mu_J=\pm\frac{\kB \bar{T}\pi}{\sqrt{3}}\sqrt{\frac{2\Delta T}{\bar{T}}}\left[1\pm\frac{\pi}{\sqrt{3}}\sqrt{\frac{2\Delta T}{\bar{T}}}\frac{D'(0)}{D(0)}\frac{\kB \bar{T}}{\Gamma}\right],
\end{equation}
which inserted into Eq.~\eqref{eq:SJ_expansion} leads to an additional correction of order $(\Delta T/\bar{T})^{3/2}$:
\begin{equation}
    \Ssh[J](0)\approx\frac{4}{h}(\kB \bar{T})^3\left[F(0)\frac{\pi^2(\pi^2-6)}{27}\frac{\Delta T}{\bar{T}}\pm K\left(\frac{\Delta T}{\bar{T}}\right)^{3/2}\right],
\end{equation}
with
\begin{equation}
    K=\frac{\kB \bar{T}}{\Gamma}\sqrt{6}\pi^3\left[F(0)\frac{2(\pi^2-6)}{81}\frac{D'(0)}{D(0)}+F'(0)\frac{\pi^2-10}{30}\right].
\end{equation}
Note that the different signs in the above equations stem from the sign of the chosen stopping voltage.

Performing similar calculations for the thermal heat noise~\eqref{eq:thermal-general}, yields
\begin{equation}
    \Sth[J](0)\approx\frac{4}{h}(k_\text{B}\bar{T})^3D\frac{2\pi^2}{3}\left(1+\frac{\Delta T}{\bar{T}}\right)
\end{equation}
for a constant transmission $D(E)=D$, and
\begin{equation}
    \begin{split}
        \Sth[J](0)&\approx\frac{4}{h}(k_\text{B}\bar{T})^3D(0)\frac{2\pi^2}{3}\left[1\pm\sqrt{\frac{3}{2}}\frac{D'(0)}{D(0)}\frac{k_\text{B}\bar{T}}{\Gamma}\sqrt{\frac{\Delta T}{\bar{T}}}\right.\\
        &\left.+\left(1+\pi^2\frac{D'(0)}{D(0)}\frac{k_\text{B}\bar{T}}{\Gamma}\right)\frac{\Delta T}{\bar{T}}\right]
    \end{split}
\end{equation}
for a weakly energy-dependent transmission.

\section{Unbounded zero-frequency heat shot noise}\label{sec:app:zero-freq-unbound}
In Ref.~\cite{Eriksson2021Sep}, it was argued that a gapped transmission can be used to achieve heat shot noise that exceeds heat thermal noise in the absence of average heat currents. Here, we show this statement in more detail.

A generic, gapped transmission function $D(E)$, e.g.~those in Figs.~\subfigref{fig:zero-freq-heat-noise}{a,d}, can be separated into two distinct energy windows as
\begin{equation}
\begin{split}
\label{eq:D_split}
D(E) &= D_<(E) + D_>(E)\\
    &\approx D_<(E)\theta(E-E_<) + D_>(E)\theta(E_>-E),
\end{split}
\end{equation}
where $\theta(x)$ is the step function and the difference $E_>-E_<$ is the energy gap.
Here, we focus on the case $\mu_\L<E_<<E_><\mu_\R$, that, as discussed in the main text, allows to achieve arbitrarily large heat shot noise.
This distinction allows us to separate the heat current according to both the transmission window and the sign of the excess energy $E-\mu_\L$.
In particular, particles contributing to the negative excess energy influx into the left contact yield the following heat flows in both energy windows
\begin{align}
    J_{\text{cool},<} =\frac2h \!\int_{-\infty}^{\mu_\L}\!dE[E-\mu_\L]D_<[f_\L - f_\R],\\
    J_{\text{cool},>} =\frac2h \!\int_{\varepsilon}^{\infty}\!dE[E-\mu_\L]D_>[f_\L - f_\R],
\end{align}
where we omitted the energy dependencies in the integrand for notational ease. The energy $\varepsilon$ is defined as the energy at which the Fermi distributions cross, namely $f_\L(\varepsilon)=f_\R(\varepsilon)$.
Similarly, particles contributing to the positive excess energy influx into the left contact yield 
\begin{align}
    J_{\text{heat},<} = \frac2h\!\int_{\mu_\L}^{E_<}\!dE[E-\mu_\L]D_<[f_\L - f_\R],\\
    J_{\text{heat},>} = \frac2h\!\int_{E_>}^{\varepsilon}\!dE[E-\mu_\L]D_>[f_\L - f_\R].
\end{align}
The zero-heat-current condition then reads
\begin{equation}
    (J_{\text{heat},<}+J_{\text{heat},>}) + (J_{\text{cool},<}+J_{\text{cool},>}) \equiv J_\text{heat} + J_\text{cool} =0.
\end{equation}
We are interested in situations in which $J_{\text{heat},>}$ is finite. Otherwise the large energy window $D_>(E)$ becomes irrelevant to the heat transport. We next consider $k_\text{B}T_\R$ to be the smallest energy scale, such that we can approximate $T_\R\approx 0$ and, consequently, $\varepsilon\approx\mu_\R$, and we consider a large gap, such that $E_>-\mu_\L\gg k_\text{B}T_\L$.
In the upper window, the Fermi functions are to good approximation given as $f_\L\approx 0$ and $f_\R\approx\theta(\mu_\R-E)$, respectively. We then have
\begin{align}
    J_{\text{cool},>} &\approx 0\\
    J_{\text{heat},>} &\approx -\frac2h\!\int_{E_>}^{\mu_\R}dE[E-\mu_\L]D_>\notag\\
    &\approx -\frac2h[E_>-\mu_\L]\!\int_{E_>}^{\mu_\R}dE D_>.
\end{align}
In the last equality, we used that at large $E_>$, $\mu_\R$ must lie very close to $E_>$ in order to satisfy the zero-heat-current condition, $J_{\text{heat}, >} = -J_\text{cool}-J_{\text{heat}, <}$.

Similarly to the heat currents, the heat shot noise is also separated according to the transmission windows, namely  $S_\text{sh}(0)^J \approx S_{\text{sh}, <}^J(0) + S_{\text{sh}, >}^J(0)$.
We now focus on the contribution generated by the upper window, which reads
\begin{align}
\label{eq:app:heat-shot-noise}
    S_{\text{sh}, >}^J(0) &= \frac4h\!\int_{E_>}^{\infty}\! dE[E-\mu_\L]^2D_>[1-D_>][f_\L-f_\R]^2,\notag\\
    &\approx \frac{4}{h} [E_>-\mu_\L]^2\!\int_{E_>}^{\mu_\R}\! dE D_>[1-D_>],\notag\\
    &\approx \frac{4}{h} [E_>-\mu_\L]^2\!\int_{E_>}^{\mu_\R}\! dE D_>,\notag\\
    &\approx 4[E_>-\mu_\L][J_\text{cool} + J_{\text{heat}, <}].   
\end{align}
Here, in the third line we recognize that, at sufficiently large gap, the energy interval $[E_>, \mu_\R]$ becomes much smaller than the scale on which $D_>$ varies.
In particular, this interval corresponds to the gap edge, where $D_>(E)$ transitions from being close to zero to being finite.
Therefore, in the energy interval $[E_>, \mu_\R]$ the transmission $D_>(E)$ is much smaller than one, $D_>(E)\ll 1$.
Since both $J_\text{cool}\approx J_{\text{cool}, <}$ and $J_{\text{heat}, <}$ do not depend on $E_>$, Eq.~\eqref{eq:app:heat-shot-noise} shows that the heat shot noise contribution of the upper transmission window grows linearly in $E_>-\mu_\L$ for large $E_>-\mu_\L$, which is the case for a large transmission gap.
Note that the specific shape of the transmission function $D(E)$ does not matter as long as it is gapped and separable as in Eq.~\eqref{eq:D_split}.

\section{Bound on zero-frequency charge and spin shot noise}\label{sec:app:zero-freq-bound}
At zero frequency, the conservation of charge and spin in the conductor demands that the shot noises measured in the left and right reservoirs are equal.
Thus, we can assume without loss of generality that the left reservoir is colder than the right one and we therefore set $T_\L<T_\R$. 
Importantly, in both cases we consider, the average spin current vanishes. Indeed, we investigate the charge noise in spin-degenerate systems, and the spin fluctuations in the absence of spin current.
Using the zero-spin-current condition 
\begin{align}
    \Sigma=\sum_\tau\int dE\,D(E)\frac{(-1)^{\delta_{\tau\downarrow}}}{4\pi}[f_{\L\tau}(E)-f_{\R\tau}(E)]=0,
\end{align}
the difference between the thermal noise and twice the thermal noise in the left reservoir can be written as
\begin{equation}\label{eq:app:zero-freq-0}
\begin{split}
    \Sth[X]-2\cT_\L^X &= \sum_\tau\frac{2x^2}{h}\!\int\!dED(E)[f_{\L\tau}(E)-f_{\R\tau}(E)]^2\\
    &+\frac{4x^2}{h}\!\int\!dED(E)\left\{f_{\R\downarrow}(E)[f_{\L\downarrow}(E)-f_{\R\downarrow}(E)] \right.\\
    &\left.+[1-f_{\R\uparrow}(E)][f_{\R\uparrow}(E)-f_{\L\uparrow}(E)]\right\},
\end{split}
\end{equation}
with $x\to\{-e,{\hbar/2}\}$ for $X\to\{I,{\Sigma}\}$. 
Here, the first integral is greater than or equal to the shot noise $\Ssh[X]$  because the integrand in Eq.~\eqref{eq:shot-general} is always positive. Moreover, it contains the additional factor $[1-D(E)]$, which is smaller than or equal to unity.
The second integral is positive due to the zero-current condition.
Indeed, calling $\varepsilon_\tau$ the energy at which $f_{\L\tau}(\varepsilon_\tau)=f_{\R\tau}(\varepsilon_\tau)$, we can split the second integral according to the sign of $[f_{\L\tau}-f_{\R\tau}]$, and use the monotonicity of $f_{\R\tau}$ to find the inequalities
\begin{equation}\label{eq:app:zero-freq-current-inequalities}
    \begin{split}
        &\!\!\!\int_{\varepsilon_\downarrow}^\infty\!\!dE  Df_{\R\downarrow}[f_{\L\downarrow}-f_{\R\downarrow}] \geq  f_{\R\downarrow}(\varepsilon_\downarrow)\int_{\varepsilon_\downarrow}^\infty\!\!dE  D[f_{\L\downarrow}-f_{\R\downarrow}],      \\
        &\!\!\!\int_{-\infty}^{\varepsilon_\downarrow}\!\!dE Df_{\R\downarrow}[f_{\L\downarrow}-f_{\R\downarrow}] \geq  f_{\R\downarrow}(\varepsilon_\downarrow)\int_{-\infty}^{\varepsilon_\downarrow}\!\!dE D[f_{\L\downarrow}-f_{\R\downarrow}].
    \end{split}
\end{equation}
Here, we omitted the energy dependence in the integrands for notational ease.
Similar inequalities are found for the last term in Eq.~\eqref{eq:app:zero-freq-0}, and, combining them, we obtain
\begin{equation}\label{eq:app:zero-freq-last}
\begin{split}
    &\frac{h}{4x^2}\left(S_\text{th}^X-2\Theta_\L^X- S_\text{sh}^X\right) \geq f_{\R\downarrow}(\varepsilon_\downarrow)\!\!\int\!\! dE D[f_{\L\downarrow}-f_{\R\downarrow}] \\
    & \qquad + [1-f_{\R\uparrow}(\varepsilon_\uparrow)]\!\!\int\!\! dE D[f_{\R\uparrow}-f_{\L\uparrow}].
\end{split}
\end{equation}
Now, if the system is spin-degenerate, namely $f_{\alpha\uparrow}=f_{\alpha\downarrow}$, the integrals in the right-hand side of \eqref{eq:app:zero-freq-last} are proportional to the charge current $I$.
Therefore, in the absence of an average charge current, namely $I=0$, the charge noise without spin imbalances satisfies the inequality~\eqref{eq:bound-shot-charge}.

Instead, if the system is spin-nondegenerate, but there is no voltage bias between the reservoirs, the occupation probabilities satisfy $f_{\alpha\uparrow}(E) = 1-f_{\alpha\downarrow}(-E)$. This electron-hole symmetry implies that $\varepsilon_\uparrow = -\varepsilon_\downarrow$, which we can use to make the right-hand side proportional to the spin current $\Sigma$.
Therefore, in the absence of an average spin current, i.e., $\Sigma=0$, the spin noise without voltage bias satisfies the inequality~\eqref{eq:bound-shot-spin}.

\section{Bound on finite-frequency charge shot noise}\label{sec:app:finite-freq-bound}
At finite frequency, the left and right autocorrelators are generally different, and we consider here the left ($\L$) charge current autocorrelation noise, i.e., we focus on the charge noise measured on the left reservoir.
Here, we prove the inequality on the symmetrized shot noise, namely Eq.~\eqref{eq:finite-freq-bound}, because it is more complex than the inequality on the absorption shot noise, namely Eq.~\eqref{eq:finite-frequency-absorption-bound}. Nonetheless, by applying the same strategies presented below, one can demonstrate the latter inequality.

The finite frequency charge shot noise and the charge thermal noise are related through
\begin{equation}\label{eq:app:shot-therl-equality}
    \Ssh[I](\omega) = \Sr[I](\omega)-\cR(\omega) -2\cT_\L(\omega) - \mathcal{K}(\omega),
\end{equation}
where $\cR(\omega) = \cR^-(\omega)+\cR^+(\omega)$ and $\cT_\L(\omega)=\cT_\L^-(\omega)+\cT^+_\L(\omega) $ are the symmetrized noise components obtained from Eqs.~\eqref{eq:thermal-R-component-absorption} and \eqref{eq:finite-freq-TL-absorption}, respectively.
Both such contributions are positive.
Moreover, the auxiliary function $\mathcal{K}(\omega)$ is defined as
\begin{equation}\label{eq:app:finite-freq-K}
    \begin{split}
        &\frac{h}{4e^2}\mathcal{K}(\omega) = \!\int\!dE D(E)[\fL-\fR]\FR\\
        &+ \!\int\!dED(E+\hbar\omega)[\FL-\FR]\fR\\
        &+\!\int\!dED(E)D(E+\hbar\omega)[\fL-\fR]\times\\
        &\qquad\qquad\times[\FL-\FR].
    \end{split}
\end{equation}
When the left reservoir is colder than the right one, namely for $T_\L<T_\R$, the first two integrals in Eq.~\eqref{eq:app:finite-freq-K}  are positive due to the zero-current condition, similarly to the case in  Eq.~\eqref{eq:app:zero-freq-last}.
However, since left and right noises are not equal, when the left reservoir is hotter ($T_\L>T_\R$) these terms are negative, thereby allowing the finite frequency charge shot noise to be larger than the finite frequency charge thermal noise. Hence, we now focus on the case $T_\L<T_\R$.
Even though the first two integrals in Eq.~\eqref{eq:app:finite-freq-K} are positive, $\mathcal{K}(\omega)$ can still be negative, for example by taking the transmission function to be $D(E)=1$ in the interval $[-E_0, E_0]$ and $D(E)=0$ elsewhere, while considering the limits $T_\L\rightarrow 0$, $T_\R\rightarrow\infty$, and choosing the frequency $\hbar\omega = E_0$. Therefore, the trivial finite frequency extension of the bound in Eq.~\eqref{eq:bound-shot-charge} does not hold.
However, a weaker version of such a bound holds also at finite frequency.
Indeed, the sum $\cT_\L(\omega)+\mathcal{K}(\omega)$ is always greater than
\begin{equation}
\cT_\L(\omega)+\mathcal{K}(\omega)\geq \frac{2e^2}{h}\!\int\!dE D(E)D(E+\hbar\omega)G_\omega(E),
\end{equation}
where the auxiliary function $G_\omega(E)$ corresponds to a combination of Fermi functions, namely
\begin{equation}
\begin{split}
        G_\omega(E) &\equiv \{\fL+\FL-2\fL\FR  \\
        &+2\fR[\FR-\FL]\}.\\
\end{split}
\end{equation}
Calling $\varepsilon$ the energy at which $f_\L(\varepsilon) = f_\R(\varepsilon)$, we consider $G_\omega(E)$ in three distinct cases according to the signs of $[\fR-\fL]$ and $[\FR-\FL]$.
In particular, when $E>\varepsilon$, $G_\omega(E)$ is bounded from below by a positive quantity
\begin{equation}
G_\omega(E)\geq \fL[1-\FR]+\FL[1-\fR].
\end{equation}
By contrast, when $\varepsilon-\hbar\omega<E<\varepsilon$, the lower bound is different but still positive,
\begin{equation}
G_\omega(E)\geq \fR[1-\fL].
\end{equation}
Finally, considering $E<\varepsilon-\hbar\omega$, we find that also in this case, $G_\omega(E)$ is greater than a positive quantity, namely
\begin{equation}
G_\omega(E)\geq \fR[1-\FL] + \FR[1-\fL].
\end{equation}
We have therefore proved that $G_\omega(E)$ is always positive, and so is therefore $\cT_\L(\omega)+\mathcal{K}(\omega)\geq 0$.
Combining this result with Eq.~\eqref{eq:app:shot-therl-equality}, we obtain the finite frequency charge shot noise bound in Eq.~\eqref{eq:finite-freq-bound}.

\bibliography{Noise_Refs.bib}

\end{document}